\documentclass[aps,showpacs,showkeys,onecolumn,floatfix,nofootinbib]{revtex4}
\usepackage{graphicx,epsfig,wrapfig,color}
\usepackage{amsmath,amssymb,wasysym}
\usepackage{hyperref}
\usepackage{color}

\newcommand{\be}{\begin{equation}}
\newcommand{\ee}{\end{equation}}
\newcommand{\ba}{\begin{eqnarray}}
\newcommand{\ea}{\end{eqnarray}}
\newcommand{\sub}[1]{\begin{subequations} #1 \end{subequations} }

\newcommand{\la}{\langle}
\newcommand{\ra}{\rangle}

\newcommand{\gs}{g_{\rm s}}
\newcommand{\gv}{g_{\rm v}}
\newcommand{\ms}{m_{\rm s}}
\newcommand{\mv}{m_{\rm v}}
\newcommand{\bs}{b_{\rm s}}
\newcommand{\bv}{b_{\rm v}}

\begin{document}
\newcommand*{\UConn}{
   Department of Physics, University of Connecticut,
   Storrs, CT 06269-3046, U.S.A.}\affiliation{\UConn}
\title{\boldmath
  Energy-momentum tensor form factor $D(t)$ of proton and neutron}
\author{Andrea Mejia}\author{Peter Schweitzer}\affiliation{\UConn}
 \date{November 2025}
\begin{abstract}
  The energy-momentum tensor (EMT) form factor $D(t)$ is
  finite and negative in hadronic models and lattice QCD when only
  strong forces are included. However, when electromagnetic forces are
  considered, the $D(t)$ of charged hadrons undergoes a dramatic change:
  at small $t$, it changes sign and diverges like $1/\sqrt{-t}$ as shown
  for the proton in the classical model by Bia\l ynicki-Birula based on
  residual nuclear forces which can be understood as a mean field approach.
  We construct an analogous neutron model and show that this framework
  accurately explains the electro\-magnetic proton-neutron mass difference.
  We demonstrate that, after appropriately rescaling the residual nuclear
  forces, the model can reproduce lattice data on the nucleon $D(t)$ up to
  $(-t)\lesssim 1\,$GeV$^2$ as well as QED effects. 
  Based on this realistic model description, we show that the proton and
  neutron $D(t)$ form factors are practically indistinguishable down to   
  $(-t) \approx 10^{-4}\rm GeV^2$ far below what can currently be accessed
  experimentally. We conclude that in the foreseeable future the $D(t)$
  form factors of proton and neutron will practically
  look the same in experiments and phenomenology.
\end{abstract}
%
%
%
%
\maketitle

\section{Introduction}
\label{Sec-1:introduction}

Hadronic EMT form factors \cite{Kobzarev:1962wt,Ji:1996ek} can be
accessed via generalized parton distribution functions in hard exclusive
reactions \cite{Mueller:1998fv,Ji:1998pc} and contain key information about
the structure of hadrons.
This work is focused on the EMT form factor $D(t)$ also known as the $D$-term
form factor \cite{Polyakov:1999gs} which has an appealing interpretation in
terms of forces inside hadrons \cite{Polyakov:2002yz,Lorce:2018egm,Freese:2021czn},
see the reviews \cite{Polyakov:2018zvc,Burkert:2023wzr,Lorce:2025oot}.
In closed systems governed by strong (short-range) forces, $D(t)$ is
negative and finite 
\cite{Ji:1997gm,Petrov:1998kf,Schweitzer:2002nm,Goeke:2007fp,Goeke:2007fq,
Wakamatsu:2007uc,Cebulla:2007ei,Jung:2013bya,Kim:2012ts,Jung:2014jja,
Mai:2012yc,Cantara:2015sna,Perevalova:2016dln,Gulamov:2015fya,
Hagler:2003jd,Shanahan:2018nnv,
Pasquini:2014vua,Hudson:2017xug,Anikin:2019kwi,
Neubelt:2019sou,Azizi:2019ytx,Gegelia:2021wnj,
Lorce:2022cle,Hackett:2023rif,Cao:2025dkv,Dehghan:2025eov,
Broniowski:2008hx}.
But when electromagnetic (long-range) forces are included then,
at small $t$, $D(t)$ changes sign and diverges like $1/\sqrt{-t}$ as
shown for charged pions in chiral perturbation theory \cite{Kubis:1999db}
and for proton in a classical model~\cite{Varma:2020crx}.
This divergence is a universal QED effect specific to all charged particles
\cite{Donoghue:2001qc,Metz:2021lqv,Freese:2022jlu} that emerges also for
non-hadronic systems \cite{Loginov:2020xoj} and for other types of long-range
forces \cite{Panteleeva:2023aiz}. The neutron of course does not exhibit such
a divergence which gives rise to important practical questions. 
Can the divergence of the proton $D(t)$ be seen experimentally? Must the proton
and neutron be treated differently in phenomenological studies of hard exclusive
reactions?

The purpose of this work is to address such questions.
For that we will, after introducing the notation
and EMT form factors (Sec.~\ref{Sec-2:EMT-FFs-and-densities}),
review the model by Bia\l ynicki-Birula (Sec.~\ref{Sec-3:model-proton})
\cite{BialynickiBirula:1993ce} explored in the aforementioned study of the
proton $D(t)$ \cite{Varma:2020crx}. We construct an analogous neutron model
(Sec.~\ref{Sec-4:model-neutron}). Among other results, an interesting highlight
is that this simple framework provides an accurate description of the electromagnetic
proton-neutron mass difference (Sec.~\ref{Sec-5:proton-vs-neutron}).
Equipped with this framework, we investigate the neutron EMT properties and compare
to proton results \cite{Varma:2020crx} (Sec.~\ref{Sec-6:EMT-FFs-and-Dterm}) 
demonstrating the usefulness of the concept of a ``regularized'' proton $D$-term
proposed in \cite{Varma:2020crx}.
We discuss how to define the neutron size which, unlike the proton case, cannot be
associated with the mean square charge radius --- but can be well defined
in terms of EMT radii (Sec.~\ref{Sec-7:radii}). 
The residual nuclear forces are not strong enough to yield a $D(t)$ as large
as seen in hadronic models or lattice QCD, but we will show that they can be
scaled up to reproduce lattice QCD data on $D(t)$ of nucleon up to
$(-t)\lesssim 1\,\rm GeV^2$ (Sec.~\ref{Sec-8:compare-to-lattice}).
Based on this realistic description of $D(t)$ of proton and neutron, which is consistent
with QED and QCD predictions, we show that the $D(t)$ of proton and neutron can
be expected to be practically indistinguishable and QED effects become noticeable
in the proton $D(t)$ only at very small $(-t)\lesssim 10^{-5}\rm GeV^2$.
Finally, we present our conclusions (Sec.~\ref{Sec-9:conclusions}).

In this work we will indirectly (in a ``reversed mode'') make use of the
interpretation of $D(t)$ and other form factors in terms of 3D spatial densities
\cite{Polyakov:2002yz} which met also criticism, see the review 
\cite{Lorce:2025oot} and references therein. However, the main points of criticism
do not apply to this work. For instance, the classical model used in this work can
be understood as a mean field approach to the description of the nucleon, a well
justified concept in the large-$N_c$ limit in QCD \cite{Witten:1979kh}.
In this limit, the nucleon mass grows as $M\propto N_c$ making relativistic recoil
corrections negligible \cite{Polyakov:2018zvc,Burkert:2023wzr,Lorce:2025oot}.
In such a situation, the criticism of exploring the Breit frame for the
interpretation of form factors as 3D densities \cite{Miller:2025zte} does not
apply, because the motion of the nucleon as a whole is non-relativistic --- while
that of its constituents can be highly relativistic \cite{Lorce:2022cle}.
Other criticism related to the notion of pressure in quantum systems \cite{Ji:2025qax}
also does not directly apply to our approach because in classical frameworks
concepts like pressure are unambiguously defined to begin with. For a more
complete overview of this aspect, we refer to \cite{Lorce:2025oot} and
\cite{Miller:2025zte,Freese:2025tqd,Ji:2025qax} for the latest developments.

\newpage
\section{EMT form factors and spatial distributions}
\label{Sec-2:EMT-FFs-and-densities}

The nucleon form factors of the total symmetric EMT operator $\hat{T}^{\mu\nu}$ 
are defined as follows \cite{Ji:1996ek,Kobzarev:1962wt}
\ba
    \la p^\prime| \hat T^{\mu\nu} |p\rangle
    = \bar u(p^\prime)\biggl[
      A(t)\,\frac{P^\mu\gamma^\nu+P^\nu\gamma^\mu}{2}
    + B(t)\ \frac{i\,(P^\mu\sigma^{\nu\rho}+P^{\nu}\sigma^{\mu\rho})\Delta_\rho}{4M}
    + D(t)\,\frac{\Delta^\mu\Delta^\nu-g^{\mu\nu}\Delta^2}{4M}\biggr]u(p)
    \label{Eq:EMT-FFs-def} \ea
where $P= \frac12(p^\prime + p)$, $\Delta = p^\prime-p$, $t=\Delta^2$. The spinors 
are normalized as $\bar u(p)\,u(p) =2M$ with spin indices omitted for brevity. 
In the Breit frame where $p^\mu=(E,-\frac12\vec{\Delta})$, 
${p^\prime}^\mu=(E,\frac12\vec{\Delta})$ and $t=-\vec{\Delta}^{\,2}$, 
the static EMT tensor is defined as
\be
    T^{\mu\nu}(\vec{r}) = \int\frac{d^3\Delta}{2E(2\pi)^3}\,
    \la p^\prime,s^\prime| \hat T^{\mu\nu} |p,s\ra\,
    e^{-i\vec{\Delta}\cdot\vec{r}}\,.
\ee
Here $T_{00}(\vec{r})$ denotes the energy density that yields the particle mass
as $M=\int d^3r\,T_{00}(\vec{r})$ and is given by \cite{Polyakov:2018zvc}
\be\label{Eq:T00-relation-FFs}
    T_{00}(r) = M  \int\!\frac{d^3\Delta}{(2\pi)^3}\,e^{-i\vec{\Delta}\cdot\vec{r}}
    \biggl(A(t)+\frac{t}{4m^2}\Bigl(B(t)-D(t)\Bigr)\biggr)
\ee
$T_{0k}(\vec{r})$ is related to the angular momentum distribution.
The spatial components describe the stress tensor 
\be
    T^{ij}(\vec{r}) = \biggl(\frac{r^i \; r^j}{r^2}-\frac{\delta^{ij}}{3}\biggr)
    s(r) + \delta^{ij}\,p(r)
\ee
where $s(r)$ is the shear force and $p(r)$ the pressure in the system
\cite{Polyakov:2018zvc} which can be determined from $D(t)$ as follows
\ba\label{Eq:def-s-p}
    s(r) = -\,\frac{1}{4M}\,r\,\frac{d\,}{dr}
    \biggl[\,\frac1r\,\frac{d\,}{dr}\,\widetilde{D}(r)\biggr]\,,
    \quad 
    p(r) = \frac{1}{6M}\,\frac{1}{r^2}\;\frac{d\,}{dr}
    \biggl[\,r^2\,\frac{d\,}{dr}\widetilde{D}(r)\biggr]\,,
    \quad
    \widetilde{D}(r) = \int\!\frac{d^3\Delta}{(2\pi)^3}
    e^{-i\vec{\Delta}\cdot\vec{r}} \, D(-\vec{\Delta}^{\,2}) \, .
\ea
Due to EMT conservation, $\nabla^i T^{ij}(\vec{r})=0$ which implies the 
following differential relation and the von Laue condition
\be
    \mbox{(a)} \quad \frac23s^\prime(r)+\frac{2}{r}s(r)+p^\prime(r)=0\,, \qquad
    \mbox{(b)} \quad \int _0^\infty \!\!dr \;r^2p(r)=0 \,.
    \label{Eq:diff-eq-von-Laue}
\ee
Mechanical stability imposes on the normal force (per unit area)
$\frac{2}{3}s(r) + p(r)$ the condition \cite{Perevalova:2016dln}
\be
    \frac{2}{3}s(r) + p(r) \ge 0 \, ,
    \label{eq:normal-force}
\ee
which holds in systems with short range forces \cite{Lorce:2025oot}.
The $D$-term $D=D(0)$ can be computed in two equivalent ways
\ba\label{Eq:D}
    D = D_{\rm press} = D_{\rm shear} \, , \quad
    D_{\rm press}   = M \int d^3 r\; r^2 p(r) \,  , \quad
    D_{\rm shear}   = - \frac{4}{15}\,  M \int d^3 r\; r^2 s(r) \, .
\ea
Two interesting quantities are the mechanical and mass mean square radii
defined as
\be\label{Eq:EMT-radii}
    \la r^2_{\rm mech}\ra = 
    \frac{\int d^3r\,r^2\bigl(\frac{2}{3}s(r) + p(r)\bigr)}
        {\int d^3r      \bigl(\frac{2}{3}s(r) + p(r)\bigr)} = 
    \frac{6D}{\int_{-\infty}^0dt\,D(t)} \, , \qquad
    \la r^2_{\rm mass}\ra = 
    \frac{\int d^3r\,r^2T_{00}(r)}
        {\int d^3r      T_{00}(r)} = 
    6A^\prime(0)-\frac{3D}{2M^2} \, . 
\ee

It is important to point out, that in  classical frameworks one deals with a
``reversed''  situation in the following sense. Form factors are the primary
properties in nature that can be determined experimentally or computed
in quantum field theoretical models, lattice QCD or other approaches.
The 3D spatial distributions are secondary properties based on 
interpretations \cite{Polyakov:2018zvc,Burkert:2023wzr,Lorce:2025oot}.
In classical models, the situation is opposite. Here the 3D spatial distributions
are the primary properties that can be directly computed, and the evaluation of
form factors in classical frameworks is based on inverting the Fourier transforms
in Eqs.~(\ref{Eq:T00-relation-FFs},~\ref{Eq:def-s-p}). One could, of course, 
wonder whether classical frameworks can be applied to describe particles.
However, the model description of the proton ``is not out of touch with reality''
\cite{BialynickiBirula:1993ce} and, as noted in Sec.~\ref{Sec-1:introduction},
this fully relativistic field theoretical framework can be viewed as a simple
model of a quantum mean field approach which allays any doubts.

The interpretation of form factors in terms of 3D distributions is justified when
the hadron size $R$ is much larger than its Compton wavelength $\lambda_C$.
For instance, this condition is well justified for nuclei, not at all for pions,
and the nucleon is a border case as assessed in \cite{Hudson:2017xug}. In the
nucleon case, however, the interpretation becomes exact in the limit of a large
number of colors~$N_c$ where baryons can be described in the mean field approach
as classical solitons of mesonic fields \cite{Witten:1979kh} and the nucleon mass
grows as $M\propto N_c$ making relativistic recoil corrections negligible
\cite{Polyakov:2018zvc,Burkert:2023wzr,Lorce:2025oot}.
In classical models, the starting point are exactly known 3D spatial distributions,
and the inversion of the Fourier transforms in
Eqs.~(\ref{Eq:T00-relation-FFs},~\ref{Eq:def-s-p}) yields reliable results for 
$(-t)\lesssim 1\,{\rm GeV}^2$ \cite{Braaten:1986iw,Ji:1991ff,Christov:1995vm}.
We will make a similar observation regarding the range of applicability for
EMT form factor computations in the present approach.

\newpage
\section{Classical proton model}
\label{Sec-3:model-proton}

In the classical model \cite{BialynickiBirula:1993ce}, the proton is made of 
dust of total mass $m_{\rm dust}$ which is bound within a spherical region of
radius $R$ by the interplay of three forces --- attractive strong force mediated
by the scalar field $\phi$ with mass $\ms$, repulsive strong force mediated 
by the vector field $V^\mu$ with mass $\mv$, and electromagnetic force 
described by the four-potential $A^\mu$ --- to which the dust particles couple 
through the pertinent coupling constants $\gs$, $\gv$, $e$.
The dynamics of the system is described by covariant field equations which 
can be found in Ref.~\cite{BialynickiBirula:1993ce} and are most conveniently
solved in the rest frame of the system where the fields are time-independent,
$A^\alpha$ and $V^\alpha$ have no spatial components, and the dust is described
by a static distribution $\rho(r)$. In this frame, the classical field equations
are given by \cite{BialynickiBirula:1993ce}
\sub{\label{Eqs:field-equations-static}\ba
        \label{eq:force-equilibrium}
        \rho \vec{\nabla}  
        ( \gv V_0 + eA_{0}-\gs \phi ) 
        & = & 0\,,\\
        \label{eq:minusdelmw}
	( - \Delta +\mv^2 ) V_0
        & = & \gv  \rho \, , \\
        \label{eq:minusdelmphi}
	( - \Delta +\ms^2 )  \phi 
        & = & \gs \rho \, , \\
        \label{eq:minusdelazero}
	- \Delta A_{0} 
        & = & e \rho \, .
        \ea}
The dust density is normalized as $\int d^{3}r \,\rho(r) = 1$.
The solutions  of Eqs.~(\ref{Eqs:field-equations-static}) are
given by \cite{BialynickiBirula:1993ce} 
(we use ``$p$'' and ``$n$'' to label model quantities and ``prot''
and ``neut'' to label physical quantities of proton and neutron)
\sub{
\label{Eqs:solution-proton}
\begin{alignat}{5}
    \label{eq:fplusfminus}
    \rho_p(r) &\;\;=\;\;& 
    \biggl(f_p^{+}(r)-f_p^{-}(r)\biggr)\,\Theta(R_p-r) 
    &\, , &
    \\
    \label{eq:eazero}
    eA_p^0(r) &\;\;=\;\;&
    e^2\biggl( 
    \frac{f_p^+(r)}{(k^+_p)^2}-\frac{f_p^-(r)}{(k^-_p)^2} 
    + \frac{2E_B^p}{e^2} \biggr) \,\Theta(R_p-r) 
    &\;\;+\;\;& 
    \frac{e^2}{4 \pi r} \;\Theta(r-R_p) \, ,
    \\
    \label{eq:gsphi}
    \gs \phi_p(r) &\;\;=\;\;&
    \gs^2 \biggl(
     \frac{f_p^+(r)}{(k^+_p)^2+\ms^2}
    -\frac{f_p^-(r)}{(k^-_p)^2+\ms^2} \biggr)\,\Theta(R_p-r) 
    &\;\;+\;\;&
    \frac{\bs^p}{4 \pi r} e^{-\ms ( r-R_p ) } \;\Theta(r-R_p) \, ,
    \\
    \label{eq:gswzero}
    \gv V_p^0(r) &\;\;=\;\;&
    \gv^2 \biggl( 
     \frac{f_p^+(r)}{(k^+_p)^2+\mv^2}
    -\frac{f_p^-(r)}{(k^-_p)^2+\mv^2} \biggr)\,\Theta(R_p-r) 
    &\;\;+\;\;& 
    \frac{\bv^p}{4 \pi r}e^{-\mv ( r-R_p ) } \;\Theta(r-R_p) \, ,
\end{alignat}
with (a misprint in the definition of $k^\pm_p$ in Eq.~(23) of 
\cite{BialynickiBirula:1993ce} was fixed in Eq.~(15)
of \cite{Varma:2020crx})
\ba
&&    f^\pm_p(r)=\frac{d_\pm^p\sin (k^\pm_p r)}{4\pi r} \, ,
    \quad 
    k^\pm_p=\sqrt{\frac{B_p\pm\sqrt{C_p}}{2Q_p^2}} \, , \nonumber\\
&&    B_p = (\gs^2-e^2) \mv^2 - ( \gv^2+e^2 ) \ms^2\, ,
    \quad 
    C_p = B^2_p - 4 e^2 Q_p^2 \ms^2 \mv^2 \,,
    \quad 
    Q_p = \sqrt{\gv^2+e^2-\gs^2} \,.
\ea}
The six constants $\bv^p,\;\bs^p,\;d_+^p,\;d_-^p,\;E_B^p,\;R_p$ can 
be fixed from the continuity and differentiability conditions of the 
fields $A_p^0(r),\;V_p^0(r), \;\phi_p(r)$ at $r=R_p$, and the 
normalization condition $\int d^3r\,\rho_p(r)=1$ can be used as
a cross check. Restoring here $\hbar$ and $c$ for convenience, the 
proton model parameters are given by \cite{BialynickiBirula:1993ce} 
\ba
	\ms c^2 = 550 \, \rm MeV, \quad 
	\mv c^2 = 783 \,\rm MeV, \quad 
	\frac{\gs^2}{\hslash c} = 91.64 \, , \quad
	\frac{\gv^2}{\hslash c} = 136.2 \, , \quad 
    \alpha = \frac{e^2}{4\pi\hslash c} = \frac{1}{137}, \, 
    \label{eq:parameters} 
\ea
The scalar and vector fields correspond to $\sigma$- and $\omega$-mesons 
with the values of $\ms$, $\mv$, $\gs$, $\gv$ taken from the QHD-I mean field 
theory model of nuclear matter \cite{nuclear-matter}. For these parameters, 
it follows 
$\bv^p=1354.13\,{\rm MeV\,fm}$,
$\bs^p=1786.38\,{\rm MeV\,fm}$,
$d_{+}^p=2.02477 /{\rm fm}^2$,
$d_{-}^p=-3.93639 /{\rm fm}^2$,
$E_B^p = -15.7089\,{\rm MeV}$,
$R_p=1.04906\,{\rm fm}$ 
from the boundary conditions of the fields at $r=R_p$.
We set $m_{\rm dust} = 953.981 \,{\rm MeV}$.\footnote{ 
    In Refs.~\cite{BialynickiBirula:1993ce,Varma:2020crx}
    $m_{\rm dust}$ was set equal to the physical proton mass 
    with the model result for the proton mass corresponding 
    to the mass of a proton bound in infinite nuclear matter 
    \cite{BialynickiBirula:1993ce}. Throughout this work,
    we choose a different scheme and identify $m_{\rm dust}$ 
    with a ``bare proton mass'' such that, after including the 
    binding energy, the sum $m_{\rm dust}+E_B^p$ reproduces the 
    physical value of the proton mass of 938.272 MeV 
    \cite{ParticleDataGroup:2024cfk}.
    This is the only difference regarding parameter fixing between
    Refs.~\cite{BialynickiBirula:1993ce,Varma:2020crx} and this work
    until Sec.~\ref{Sec-8:compare-to-lattice}. 
    \label{footnote-parameter-fixing}}

The proton mass is $M_{\rm prot}=m_{\rm dust}+E_B^p$ where the binding energy 
$E_B^p = -15.71\,{\rm MeV}$ \cite{BialynickiBirula:1993ce} is to be
compared to the bulk binding energy per nucleon in nuclear matter 
of $-15.75\,{\rm MeV}$ \cite{nuclear-matter}. Since the dust in the
proton model is bound by same nuclear forces that bind nuclei in 
nuclear matter, this is an important consistency check of the model 
\cite{BialynickiBirula:1993ce}. The mean square dust radius given  
by $\la r^2_{\rm dust}\ra = \int d^3r\,r^2\rho(r)$ yields the proton 
charge radius whose numerical value $0.710\,{\rm fm}$ in the model
underestimates the experimental result by $15\,\%$ but is 
``not completely out of touch with reality'' 
\cite{BialynickiBirula:1993ce} making this model sufficient for
the purposes of Refs.~\cite{BialynickiBirula:1993ce,Varma:2020crx}
(and this work until Sec.~\ref{Sec-8:compare-to-lattice}).

\section{Classical neutron model}
\label{Sec-4:model-neutron}

We proceed with the construction of the analogous neutron model by
keeping everything exactly as in the proton case, except for ``removing''
the electromagnetic interaction. In this way, the only difference between
proton and neutron is due to electromagnetic effects. We make no effort
to attempt considering isospin breaking effects. 

Notice that by taking the limit $e\to 0$ in Eqs.~(\ref{Eqs:field-equations-static}) 
the dust becomes electrically neutral. This implies that the neutron electric
form factor $G_{En}(t)$ vanishes. Experimentally, $G_{En}(t)$ is found much
smaller than the proton electric form factor $G_{Ep}(t)$.
Numerically, $G_{En}(t)/G_{Ep}(t)\lesssim 20\%$ in the region
$0 \le (-t) \lesssim 1\,\rm GeV^2$ \cite{Punjabi:2015bba}
(which, as we shall see, corresponds to the range of applicability of this approach).
To the extent that this number can be considered numerically small, one may argue
that we neglect a small effect. More importantly, with the main focus of this work
on mechanical (rather than electrical) properties, this simple neutron model is
sufficient.

In the limit $e\to0$, the Poisson equation (\ref{eq:minusdelazero}) 
becomes a Laplace equation with the simple solution $A_0(r)=0$. 
The static field equations (\ref{Eqs:field-equations-static}a-d) 
are replaced in the neutron case by
\sub{\label{neut-Eqs:field-equations-static}\ba
        \label{neut-eq:force-equilibrium}
        \rho \vec{\nabla}  
        ( \gv V_0 - \gs \phi) 
        & = & 0\,,\\
        \label{neut-eq:minusdelmw}
	( - \Delta +\mv^2 ) V_0
        & = & \gv  \rho \, , \\
        \label{neut-eq:minusdelmphi}
	( - \Delta +\ms^2 )  \phi 
        & = & \gs \rho \, 
\ea}
with $\int d^3r\;\rho(r) = 1$ and $\rho(r)=0$ for $r>R_n$ as in the 
proton case. The solution of (\ref{neut-Eqs:field-equations-static}) 
can be derived from the proton solution (\ref{Eqs:solution-proton}) 
by taking the limit $e\to 0$. In this limit \sub{
\be
    \lim_{e\to0} B_p = \lim_{e\to0} \sqrt{C_p} 
    = \gs^2 \mv^2 - \gv^2 \ms^2 \equiv B_n\,,
    \quad 
    \lim_{e\to0} Q_p = \sqrt{\gv^2 -\gs^2} \equiv Q_n \,.
\ee
For $e\to0$ the constant $k^+_p$ has a finite limit, but $k^-_p$ 
vanishes according to
\be\label{Eq:kpm-neut}
    \lim_{e\to0} k^+_p = k^+_n = \frac{\sqrt{B_n}}{Q_n}\; , 
    \quad
    k^-_p = e\;\frac{\ms \mv}{\sqrt{B_n}}
    +{\cal O}(e^2) \;\; \to \;\; 0\,.
\ee
The latter implies that in Eq.~(\ref{eq:eazero}) the limit $e\to0$ must be 
taken with care. This equation yields the constraint
\be\label{Eq:constraint}
      \lim_{e\to0} eA_p^0(r) 
    = \lim_{e\to0} \biggl( 2E^p_B - \frac{e^2f_p^-(r)}{(k^-_p)^2} \biggr) 
    = 2E^n_B - \frac{\sqrt{B_n}}{\ms \mv}\;\lim_{e\to0} \frac{e d_-^p}{4\pi} 
    \stackrel{!}{=} 0.
\ee
Since the dust in the neutron must be bound, we expect $E^n_B \neq 0$. This implies
that in Eq.~(\ref{Eq:constraint}) the constant $d_-^p$ must diverge as $1/e$.
The divergent constant $d_-^p$ appears always inside $f^-_p(r)$ in combination with
$\sin(k^-_p r)$. In this combination the divergence of $d_-^p\propto 1/e$ and the
vanishing of $k^-_p\propto e$ compensate each other such that 
\be
    \lim_{e\to0} f^-_p(r)
    = \lim_{e\to0} \frac{d_-^p\sin (k^-_p r)}{4\pi r} 
    = \lim_{e\to0} \frac{d_-^pk^-_p}{4\pi}
    = \frac{2\ms^2 \mv^2}{B_n} \;E_B^n
    \equiv f^-_n(r)
    \, .
\ee}
In other words, in the neutron case the function $f^-_n(r)$
becomes a constant, while $f^+_n(r)$ retains the same form 
as in the proton case, albeit with different constants. 

Summarizing these findings in a notation chosen
to maximally resemble the notation of the proton solutions
(\ref{Eqs:solution-proton}), we obtain the neutron solutions 
(labeled by ``$n$'' for the neutron) given by
\sub{\label{Eqs:solution-neutron}
\begin{alignat}{5}
    \label{neut-eq:fplusfminus}
    \rho_n(r) &\;\;=\;\;& 
    \biggl(f^n_{+}(r)-f^n_{-}(r)\biggr)\,\Theta(R_n-r) 
    &\, , &
    \\
    \label{neut-eq:gsphi}
    \gs \phi_n(r) &\;\;=\;\;&
    \gs^2 \biggl(
     \frac{f_{+}^n(r)}{(k^+_n)^2+\ms^2}
    -\frac{f_{-}^n(r)}{(k^-_n)^2+\ms^2} \biggr)\,\Theta(R_n-r) 
    &\;\;+\;\;&
    \frac{\bs^n}{4 \pi r} e^{-\ms ( r-R_n ) } \;\Theta(r-R_n) \, ,
    \\
    \label{neut-eq:gswzero}
    \gv V_0^n(r) &\;\;=\;\;&
    \gv^2 \biggl( 
     \frac{f_{+}^n(r)}{(k^+_n)^2+\mv^2}
    -\frac{f_{-}^n(r)}{(k^-_n)^2+\mv^2} \biggr)\,\Theta(R_n-r) 
    &\;\;+\;\;& 
    \frac{\bv^n}{4 \pi r}e^{-\mv ( r-R_n ) }\;\Theta(r-R_n) \, ,
\end{alignat}
with the functions $f^\pm_n(r)$ and constants $k^\pm_n$ given by 
\ba
&&  f^+_n(r)=\frac{d_+^n\sin(k^+_n r)}{4\pi r}\,,
    \quad 
    f^-_n(r)=\frac{2\ms^2 \mv^2}{B_n^2}\,E_B^n\,,
    \quad
    k^+_n=\frac{\sqrt{B_n}}{Q_n}\,, \nonumber\\
&&  B_n = \gs^2\mv^2 - \gv^2\ms^2 \,,
    \quad 
    Q_n = \sqrt{\gv^2-\gs^2} \,.
\ea}
The five constants $\bv^n,\;\bs^n,\;d_+^n,\;E_B^n,\;R_n$ are 
fixed by the continuity and differentiability conditions of the 
scalar and vector fields at $r=R_n$ and the normalization 
$\int d^3r\,\rho_n(r)=1$. With the choice of parameters 
as in the proton case in Eq.~(\ref{eq:parameters}) 
(except that $e$ is set to zero) the constants assume the values
$\bv^n=1381.63\,{\rm MeV\,fm}$,
$\bs^n=1816.96\,{\rm MeV\,fm}$,
$d_+^n=2.06438\,{\rm fm}^2$,
$E_B^n = -16.6614\,{\rm MeV}$,
$R_n=1.03958\,{\rm fm}$.
We discuss the results of this model in the 
next section.

\section{Neutron and proton EMT distributions}
\label{Sec-5:proton-vs-neutron}

\begin{figure}[b!]
  \begin{center}
    \includegraphics[height=3.5cm]{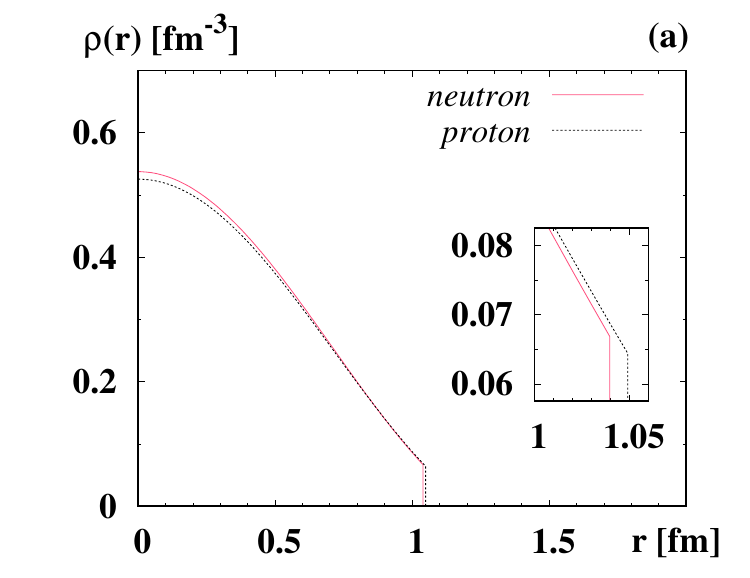}%
    \includegraphics[height=3.5cm]{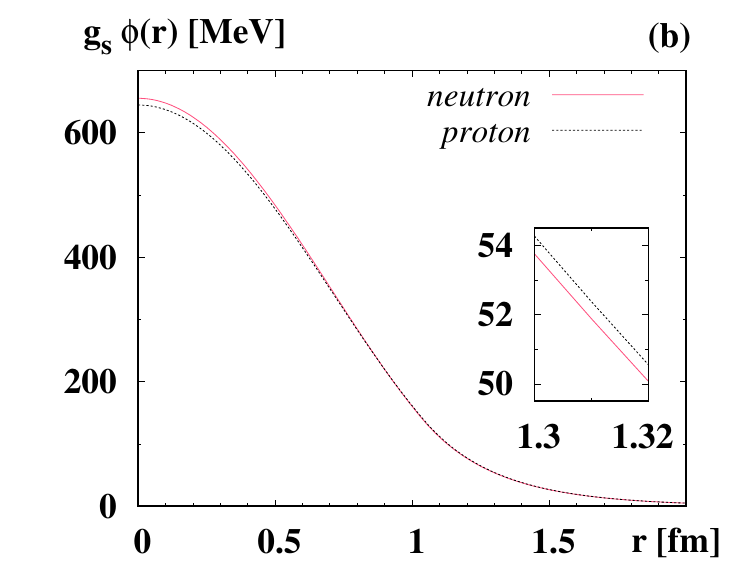}%
    \includegraphics[height=3.5cm]{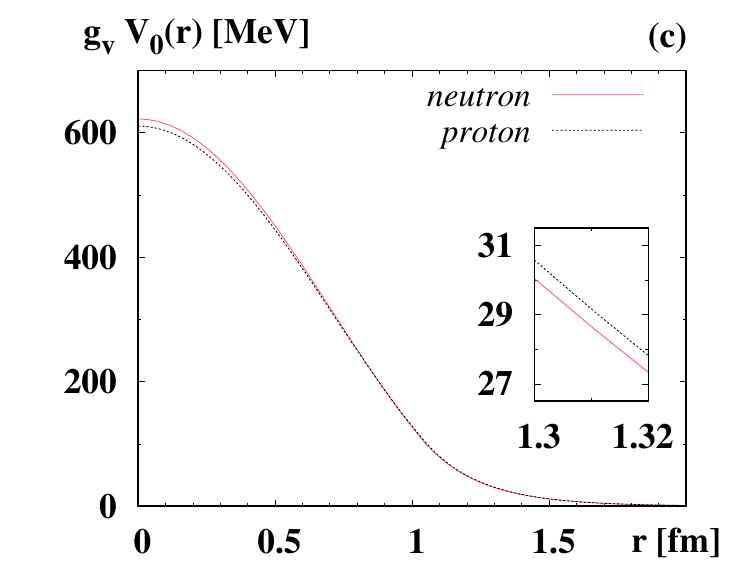}%
    \includegraphics[height=3.5cm]{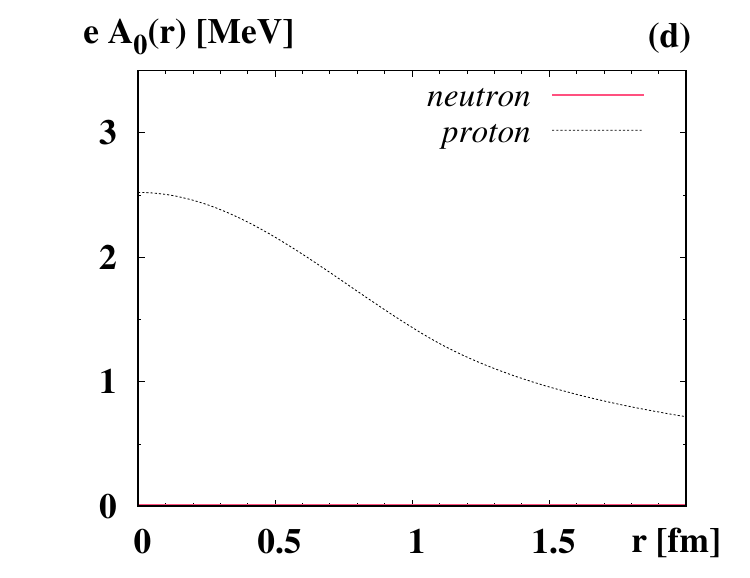}%
  \end{center}
  \caption{\label{fig1}
    Dust distribution $\rho(r)$,
    scalar potential $\gs\phi(r)$,
    vector potential $\gv V_0(r)$, and
    Coulomb potential $eA_0(r)$ as functions of $r$
    in the neutron (this work) and proton \cite{Varma:2020crx}.
    The inserts illustrate that the fields are more strongly localized 
    in the neutron. } 
\end{figure}

In Fig.~\ref{fig1} we show the neutron results in comparison to the proton
results from Ref.~\cite{Varma:2020crx} for the dust distribution $\rho(r)$
and the potentials multiplied by the respective coupling constants,
$\gs\phi(r)$, $\gv V_0(r)$, $eA_0(r)$. Notice that $R_n = 1.04\,{\rm fm}$ vs
$R_p = 1.05\,{\rm fm}$ which implies that the neutron is slightly smaller
than the proton (we will discuss the question how to define the neutron size
in detail in Sec.~\ref{Sec-7:radii}). We observe in Fig.~\ref{fig1}
that $\rho (r)$, $\gs\phi(r)$, $\gv V_0(r)$ of the neutron are slightly larger
in the inner region up to $r\lesssim 0.9\,\rm fm$ than the corresponding proton
functions. Around $r\approx 0.9\,\rm fm$ the picture flips and the proton functions 
are slightly larger as illustrated by the inserts in Figs.~\ref{fig1}a-c. 
The Coulomb potential is much smaller than the strong fields in the proton
and identically zero in the neutron --- which makes the difference.
The electrostatic repulsion of the charged dust leads to a ``swelling'' of the
proton as compared to the neutron. The dust distribution is therefore more compact
in the neutron. Since the dust is also the source of the strong fields, the latter
are slightly more strongly localized in the neutron.

\begin{figure}[t!]
  \begin{center}
    \includegraphics[height=3.5cm]{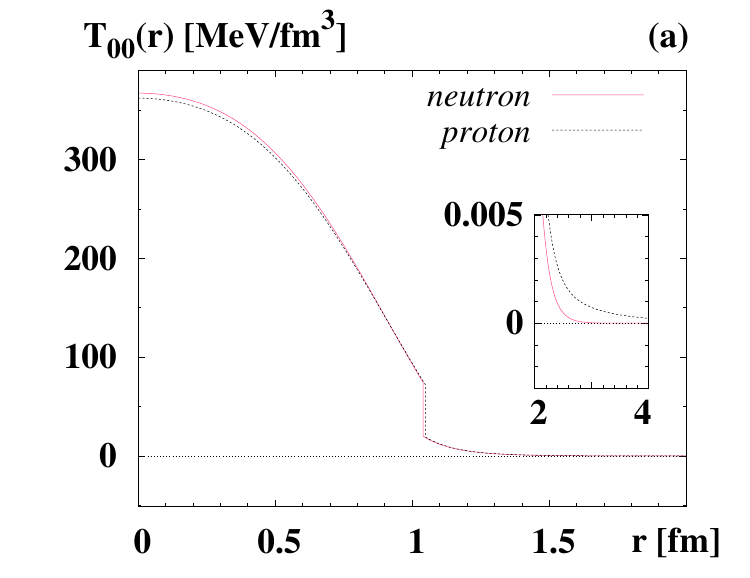}%
    \includegraphics[height=3.5cm]{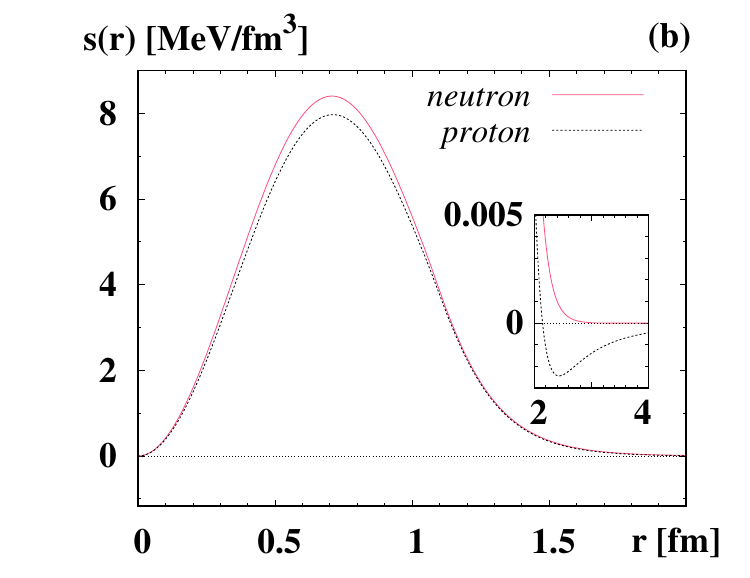}%
    \includegraphics[height=3.5cm]{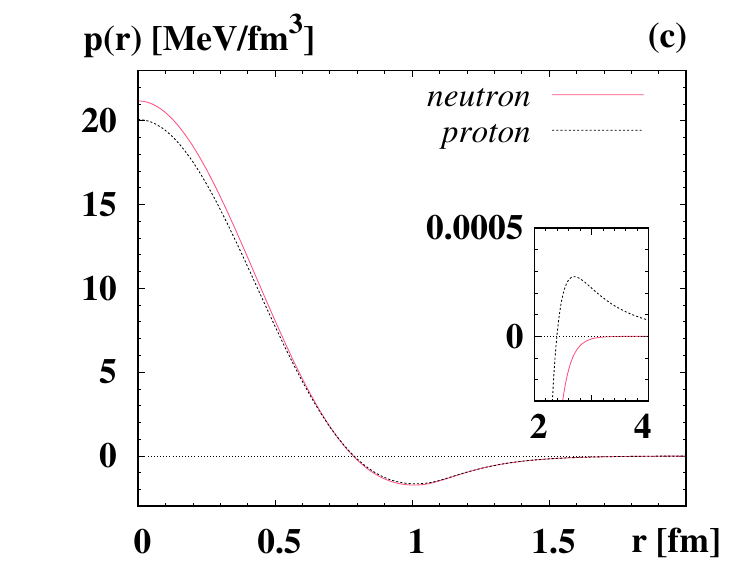}%
    \includegraphics[height=3.5cm]{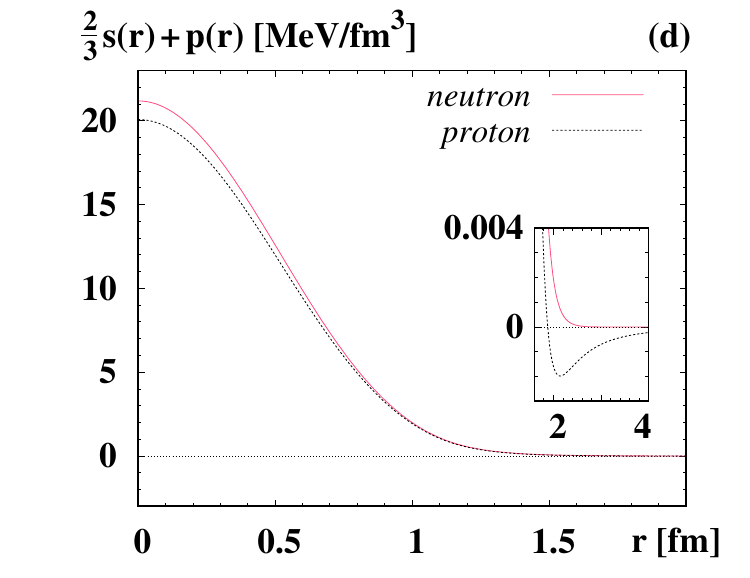}
  \end{center}
  \caption{\label{fig2}
    EMT densities $T_{00}(r)$, $s(r)$, $p(r)$ and normal force per unit area
    $\frac23s(r)+p(r)$ as functions of $r$ for neutron (this work) and proton
    \cite{Varma:2020crx}. The inserts show the outer regions $r\gtrsim 2\,\rm fm$
    where the QED effects are manifest.}	
\end{figure}

Next we discuss the EMT densities $T_{00}(r)$, $p(r)$, $s(r)$ whose model
expressions are given by \cite{BialynickiBirula:1993ce,Varma:2020crx}
\sub{\ba
    T_{00}(r) = T_{00}(r)_{\rm strong} +  T_{00}(r)_{\rm em} \, , \qquad
    p(r) =           p(r)_{\rm strong} +       p(r)_{\rm em} \, , \qquad
    s(r) =           s(r)_{\rm strong} +       s(r)_{\rm em} \, , \qquad
\label{eq:s}
\ea
where
\ba
    T_{00}(r)_{\rm strong} &=& 
        + \frac{1}{2} \phi^\prime(r)^2
        + \frac{1}{2} V_0^\prime(r)^2
        + \frac{1}{2} \ms^2 \phi(r)^2
        + \frac{1}{2} \mv^2 V_0(r)^2
        + \Bigr(m_{\rm dust}-\gs \phi(r) \Bigl)  \rho(r) , 
        \label{eq:T00-strong}\\
    p(r)_{\rm strong} &=& 
        -\frac{1}{6} \phi^\prime(r)^2
        +\frac{1}{6}V_0^\prime(r)^2
        -\frac{1}{2}\ms^2\phi(r)^2
        +\frac{1}{2}\mv^2V_0(r)^2 \,, 
        \label{eq:p-strong}\\
    s(r)_{\rm strong} &=& 
        \phantom{-\frac{1}{6} }   \phi^\prime(r)^2
        \phantom{ \frac{1}{6} } - V_0^\prime(r)^2 \,, \phantom{\frac11}
        \label{eq:s-strong}  \\
    T_{00}(r)_{\rm em} &=& \phantom{-} \frac12 A_0^\prime(r)^2 \,,\label{eq:T00-em}\\
    p(r)_{\rm em} &=& \phantom{-} \frac{1}{6}  A_0^\prime(r)^2 \,,\label{eq:p-em}\\
    s(r)_{\rm em} &=& \hspace{3mm}           - A_0^\prime(r)^2 \,, \phantom{\frac11}\label{eq:s-em}
\ea}
with the understanding that these expressions are to be evaluated 
with respectively $\rho_p(r)$, $\phi_p(r)$, $V_p^0(r)$, $A_p^0(r)$ 
in Eqs.~(\ref{Eqs:solution-proton}a-e) in the proton case, and with
$\rho_n(r)$, $\phi_n(r)$, $V_0^n(r)$ in Eqs.~(\ref{Eqs:solution-neutron}a-d) 
and the Coulomb potential set to zero in the neutron case.
In Figs.~\ref{fig2}a-d we show the EMT distributions as functions of $r$ 
for neutron in comparison to the proton results \cite{Varma:2020crx}. 
In the region $r\lesssim 2\,\rm fm$, the proton and neutron distributions look
much alike up to about $1\,\rm fm$. The neutron exhibits a slightly larger energy
density and slightly stronger forces, appearing more compact compared to the slightly
more spread out proton. These effects are small because in the inner region where
the dust is located, the Coulomb force is overwhelmed by the strong forces.

The picture is crucially different in the outer region $r\gtrsim 2\,{\rm fm}$ 
illustrated by the inserts in Figs.~\ref{fig2}a-d. Here, the neutron distributions 
exhibit exponential decays governed by the Yukawa tail of the scalar field 
(the more massive vector field decays faster) while the proton densities 
exhibit power decays dictated by the Coulomb field, see Table~\ref{tab1}.
Overall, in the neutron: 
(i) $s(r)$ is always positive,
(ii) $p(r)$ remains negative after exhibiting a single node, 
(iii) $\frac23\,s(r)+p(r)$ is always positive.
These are common features found for ground state solutions of systems governed by 
short-range forces \cite{Polyakov:2018zvc}. These features are also observed in 
the proton up to distances $r\lesssim 2\,\rm fm$. Beyond that point, however,
the Coulomb contribution begins to dominate the proton asymptotics, and:
(i) causes $s(r)$ to turn negative (insert of Fig.~\ref{fig2}b),
(ii) generates a second node in $p(r)$ turning it positive at large $r$ (insert of Fig.~\ref{fig2}c),
(iii) makes $\frac23\,s(r)+p(r)$ become negative (insert of Fig.~\ref{fig2}d).
In short, the effect of the long-range Coulomb field is to reverse signs of 
$s(r)$, $p(r)$ and the normal force as observed in \cite{Varma:2020crx}.
This large-distance behavior of the EMT distributions is a generic QED effect
arising from one-loop QED corrections
\cite{Donoghue:2001qc,Varma:2020crx,Metz:2021lqv,Freese:2022jlu}.

\begin{table}[b!]
\begin{tabular}{c|rl|rl|rl|rl}
  \hspace{2cm} 
& \hspace{1cm} & \hspace{2cm} 
& \hspace{1cm} & \hspace{2cm} 
& \hspace{1cm} & \hspace{2cm} 
& \hspace{1cm} & \hspace{2cm} \\ 
  && $T_{00}(r)$ && $s(r)$  && $p(r)$    &&  $\hspace{-5mm}\frac23s(r)+p(r)$ \\
&&&&&&&&\vspace{-2mm} \\ \hline \vspace{-3mm} 
&&&&&&&&\\
neutron      & $            c_n$&$\displaystyle\frac{\ms^2  }{r^2}\,e^{-2\ms r}$ 
             & $            c_n$&$\displaystyle\frac{\ms^2  }{r^2}\,e^{-2\ms r}$
             & $  -\frac23\,c_n$&$\displaystyle\frac{\ms^2  }{r^2}\,e^{-2\ms r}$
             & $            c_n$&$\displaystyle\frac{\ms^{ }}{r^3}\,e^{-2\ms r}$ \\ 
&&&&&&&&\vspace{-2mm}\\
proton       & $   \frac12\,c_p$&$\displaystyle\frac{1}{r^4}$ 
             & $         -\,c_p$&$\displaystyle\frac{1}{r^4}$  
             & $   \frac16\,c_p$&$\displaystyle\frac{1}{r^4}$
             & $-\,\frac12\,c_p$&$\displaystyle\frac{1}{r^4}$\vspace{-3mm}\\
&&&&&&&&             
\end{tabular}
\caption{\label{tab1} 
Behavior of neutron and proton EMT distributions at asymptotic
distances $\ms r\gg1$. The neutron asymptotics are determined 
by the scalar field with $c_n = b_n^2 e^{2\ms R_n}/(4\pi \gs)^2$
and the proton ones by the Coulomb field with 
$c_p = \frac{\alpha}{4\pi}\,\hbar c$.}
\end{table}

To demonstrate the theoretical consistency of the approach, let us mention
that the differential equation (\ref{Eq:diff-eq-von-Laue}a) holds in the
neutron and proton models, and forces in the von Laue condition
(\ref{Eq:diff-eq-von-Laue}b) balance each other exactly as follows
\ba
    \mbox{neutron:} &&
    \int_0^\infty dr\;r^2p_i(r) = \begin{cases} 
              -11.092\,{\rm MeV} & \quad {\rm for} \quad i = {\rm scalar}, \\
    \phantom{-}11.092\,{\rm MeV} & \quad {\rm for} \quad i = {\rm vector},  \end{cases}       
    \nonumber\\
    \mbox{proton:} && 
    \int_0^\infty dr\;r^2p_i(r) = \begin{cases} 
               -10.916\,{\rm MeV} & \quad {\rm for} \quad i = {\rm scalar}, \\
     \phantom{-}10.891\,{\rm MeV} & \quad {\rm for} \quad i = {\rm vector}, \\
     \phantom{-1}0.025\,{\rm MeV} & \quad {\rm for} \quad i = {\rm Coulomb}, 
     \end{cases}       
\ea
which reinforces what we saw in Fig.~\ref{fig2}: in the neutron the forces
are slightly larger than in the proton. In the neutron, the normal forces
(\ref{eq:normal-force}) are always positive. In the proton, they are
positive up $1.88\,\rm fm$, see Fig.~\ref{fig2}d and Table~\ref{tab1}.
As there is no dust at that distance, this does not impact the
mechanical stability of the proton \cite{Varma:2020crx}. 

Finally, we discuss the proton-neutron mass difference. We remind that in this
model isospin violating effects are not considered and the proton-neutron mass
difference is entirely due to electromagnetic effects. We obtain a very good agreement
with a lattice QCD study where QED effects were considered \cite{BMW:2014pzb}
\ba\label{Eq:mass-comparison}
    \mbox{classical model:} && 
    (M_{\rm prot} - M_{\rm neut})_{\rm e.m.} = (E_B^p-E_B^n) = 0.95\,\rm MeV\,, \\
    \mbox{lattice QCD+QED:} &&  
    (M_{\rm prot} - M_{\rm neut})_{\rm e.m.} = (1.00\pm 0.07_{\rm stat} \pm 0.14_{\rm syst})
    \,\rm MeV,
    \nonumber
\ea
The fact that electromagnetic effects make the proton mass larger compared
to the neutron mass can also be intuitively understood in the classical model. 
One can think of ``assembling'' the proton solution from a neutron solution
by moving electric charges from infinity to a finite volume within the radius 
$R_p$ which requires a work $W_{\rm el} = (M_p - M_n)_{\rm e.m.} = 0.95\,\rm MeV$ 
to overcome the electrostatic repulsion which includes a slight rearrangement 
of the dust distribution (this ``thought experiment'' assumes that the dust 
particles can carry any continuous amount of electric charge and there is 
no confinement which is the case \cite{BialynickiBirula:1993ce} in the model).
In nature, isospin breaking effects due to the current quark masses 
$m_d > m_u$ make a contribution that is larger and of opposite sign 
to Eq.~(\ref{Eq:mass-comparison}) \cite{BMW:2014pzb} which explains
the experimental observation
$(M_n-M_p)_{\rm exp} = 1.293 \,{\rm MeV}$
\cite{ParticleDataGroup:2024cfk}.

\section{\boldmath EMT form factors and the $D$-term}
\label{Sec-6:EMT-FFs-and-Dterm}

\begin{figure}[b!]
  \begin{center}
    \includegraphics[height=3.6cm]{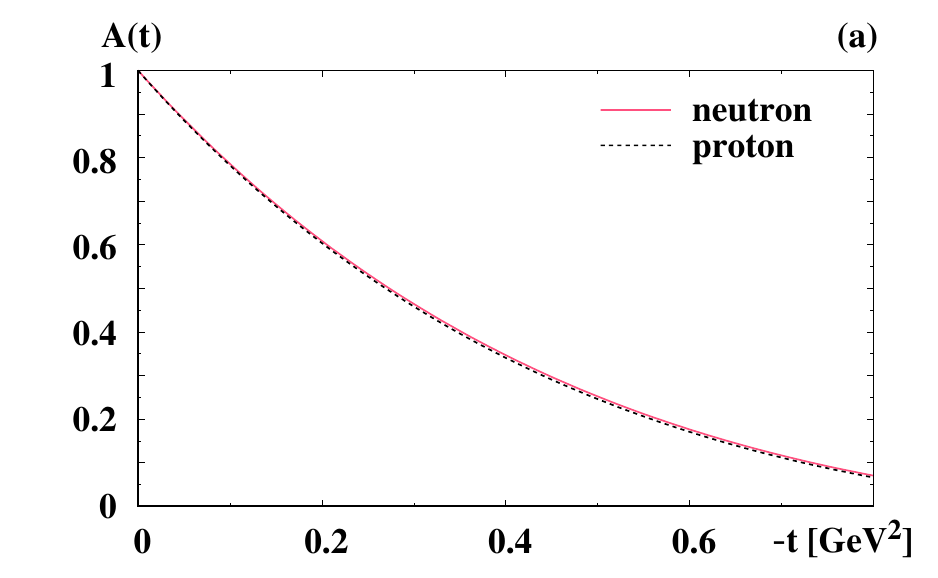}
    \includegraphics[height=3.6cm]{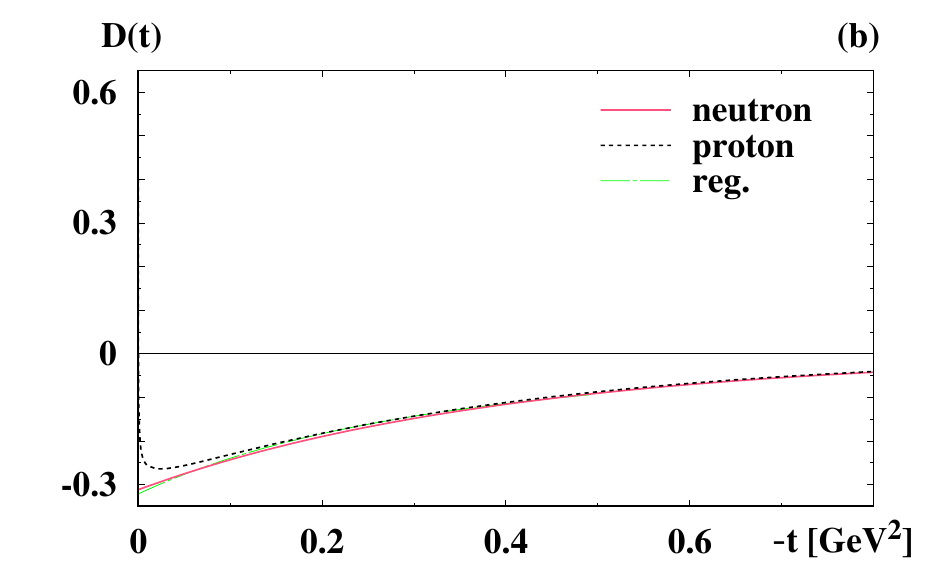}
    \includegraphics[height=3.6cm]{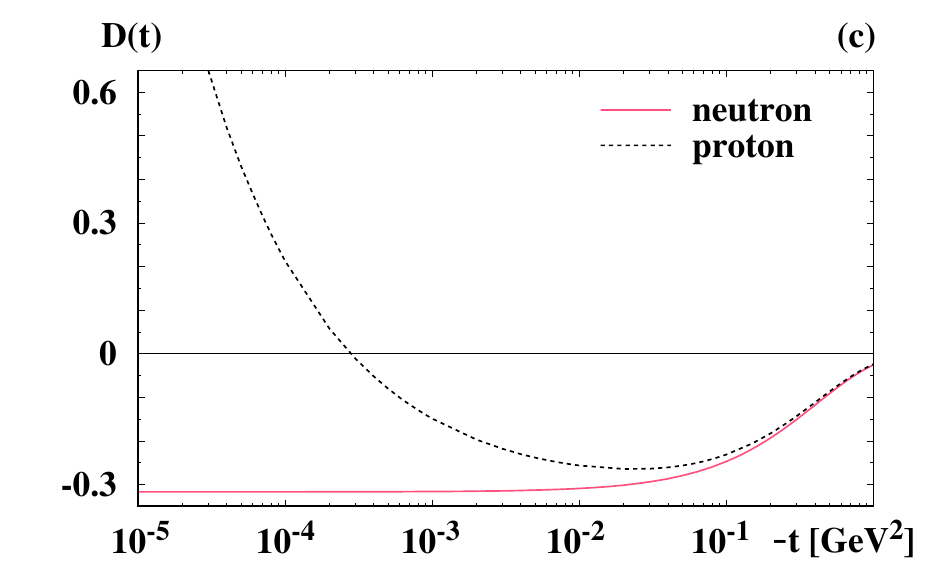}
  \end{center}
  \caption{\label{fig3-A-D-FFs}
    Form factors $A(t)$ (a) and $D(t)$ (b) of neutron and proton vs.\ $(-t)$
    in the classical model. (c) The same as panel (b) but with logarithmic 
    $t$-scale to display the region $(-t)\ll 0.1\,\rm GeV^2$ where neutron and proton 
    $D(t)$ exhibit different properties.}
\end{figure}

The proton $D(t)$ was computed in
\cite{Varma:2020crx} from $p(r)$ and $s(r)$ in (\ref{Eq:def-s-p}) which yields
the same result. Here we use the same technique for the neutron case.
The form factor $A(t)$ was not discussed in \cite{Varma:2020crx}. We compute it using 
Eq.~(\ref{Eq:T00-relation-FFs}) where, strictly speaking, the knowledge of the form factor 
$B(t)=A(t)-2J(t)$ is needed with $J(t)$ the angular momentum form factor satisfying 
$J(0)=\frac12$ for a spin-$\frac12$ particle. $J(t)$ can be determined from the 
$T_{0k}(\vec{r})$ components of the EMT which vanish for the static (non-rotating)
proton and neutron solutions in the classical model. In principle, one could use standard 
projection techniques \cite{Rajamaran} to assign definite spin quantum numbers to our 
``mean field'' solutions to obtain non-zero results for $T_{0k}(\vec{r})$.
We will refrain from this step and content ourselves with a very good approximation which
consists in neglecting $B(t)$ in Eq.~(\ref{Eq:T00-relation-FFs}). This step is justified
in the large $N_c$ limit $|t|\sim N_c^0 \ll M^2\sim N_c^2$ 
where the form factors $A(t)$, $B(t)$, $J(t)$ behave like $N_c^0$ while $D(t)\sim N_c^2$ 
\cite{Polyakov:2018zvc} which allows one to neglect $|B(t)|\ll |D(t)|$ in 
Eq.~(\ref{Eq:T00-relation-FFs}). This result is strongly supported by lattice QCD results 
where $|B(t)|\ll |D(t)|$ is found \cite{Hackett:2023rif}. The form factors $A(t)$ and $D(t)$ 
of proton and neutron obtained in this way in the classical model are shown in Fig.~\ref{fig3-A-D-FFs}.

The form factor $A(t)$ must satisfy the constraint $A(0)=1$, and the model results
comply with this requirement. Numerically, the proton and neutron $A(t)$ form factors 
can hardly be distinguished on the scale of Fig.~\ref{fig3-A-D-FFs}a. In Fig.~\ref{fig3-A-D-FFs}b we 
show $D(t)$ of the neutron and proton using the same scale as in Fig.~\ref{fig3-A-D-FFs}a to
demonstrate that also in the case of $D(t)$ the proton and neutron results can 
hardly be distinguished on the scale of the figure for $(-t)\gtrsim 0.1\,\rm GeV^2$. 
The difference between neutron and proton $D(t)$ becomes apparent only in the region 
$(-t)\ll 0.1\,\rm GeV^2$ displayed in Fig.~\ref{fig3-A-D-FFs}c with a logarithmic scale 
for the $(-t)$-axis for better visibility. As can be seen in Fig.~\ref{fig3-A-D-FFs}c, the neutron 
$D(t)$ form factor remains negative and approaches a well-defined value for $t\to 0$. 
The proton $D(t)$ form factor starts to deviate from the neutron case more and more strongly 
with decreasing $(-t)$ and changes sign at $t\simeq -3\times10^{-4}\,\rm GeV^2$. 
Eventually for $(-t) \lesssim 10^{-6}\,{\rm GeV}^2$ the proton $D(t)$ diverges 
according to the QED prediction \cite{Donoghue:2001qc}
\be\label{eq:D-term-QED}
      D(t) = \alpha\,\frac{\pi M}{4\sqrt{-t}} + \dots \,,
\ee
where the dots indicate subleading terms as $t\to0$. In \cite{Varma:2020crx} the proton
$D(t)$ form factor was studied down to $(-t) = 10^{-8}\,{\rm GeV}^2$ and the classical 
proton model was demonstrated to reproduce the QED prediction (\ref{eq:D-term-QED}).

In the model, the reason for the divergence of the proton $D(t)$ in
Eq.~(\ref{eq:D-term-QED}) is due to the large-$r$ behavior of the Coulomb field.
The neutron EMT distributions decay exponentially at large $r$,
see Table~\ref{tab1}, and the integrals defining the $D$-term in Eq.~(\ref{Eq:D}) converge 
such that one obtains a well-defined and unambiguous result for $D_{\rm neut}$. In the proton 
case, however, the Coulomb field determines the asymptotics, see Table~\ref{tab1}, and
the corresponding integrals diverge. In Ref.~\cite{Varma:2020crx} it is was proposed 
to introduce a ``regularized'' proton $D$-term $D_{\rm prot, reg}$.
In the remainder of this section, we will review how this regularization works and, based on
the neutron results, show that it is reasonable.
In Sec.~\ref{Sec-8:compare-to-lattice} we will provide an argument as to why a regularized proton
$D$-term makes sense also from a practical point of view.

The starting point to define a regularized $D_{\rm prot, reg}$ in \cite{Varma:2020crx} 
was the observation that, in a theory where the integrals converge, one may use 
$D_{\rm press}$ or $D_{\rm shear}$ in Eq.~(\ref{Eq:D}), or any linear combination 
$D(\zeta)=\zeta D_{\rm press}+(1-\zeta)D_{\rm shear}$ to obtain the same result for 
any $\zeta$. In the proton case, $D(\zeta)$ diverges for all $\zeta$ except for 
one special value $\zeta_{\rm reg}=\frac83$ when
\be\label{Eq:D-prot-reg}
    D_{\rm prot, reg} = D(\zeta)\bigl|_{\zeta=\frac83}= M \int d^3 r\; r^2 
    \biggl[\tfrac83\,p(r) + \tfrac49\,s(r)\biggr]\,
\ee
is finite because the square bracket under the integral corresponds to 
$\frac49\,[6p(r)+s(r)]$ where the Coulomb tails in Table~\ref{tab1} 
cancel out exactly \cite{Varma:2020crx}. 

Let us present now an independent regularization method that can be
applied to the $D$-term and other quantities. Let $F(r)$ be an EMT 
distribution in the proton with the asymptotics in Table~\ref{tab1}. 
Consider the integral $I_n = \int d^3r\,r^n F(r)_{\rm reg}$ with $n>1$ 
which diverges at large $r$ and needs to be regularized as indicated by 
the subscript. We subtract under the integral of $I_n$ a second term which 
is identically zero due to Eq.~(\ref{Eq:diff-eq-von-Laue}a) as follows
\ba
    I_n = \int d^3r\,\biggl[r^n F(r)
    - Z_F\,r^{n+1}
    \Bigl\{\tfrac23\,s'(r)+\tfrac2r\,s(r)+p'(r)\Bigr\}\biggr]_{\rm reg} 
    \nonumber
\ea
with the constant $Z_F$ to be defined shortly. In the next step, we carry 
out an integration by parts in the second term and ``neglect'' a divergent boundary 
term at $r\to\infty$ which we define to be permissible under the regularization 
prescription. We obtain 
\ba
    I_n = \int d^3r\,r^n \biggl[F(r) 
    + Z_F\Bigl\{\tfrac23\,n\,s(r)+(n+3)\,p(r)\Bigr\}\biggr]_{\rm reg} 
    \nonumber
\ea
At large $r$ the linear combination
$\tfrac23\,n\,s(r)+(n+3)\,p(r) = -\,\tfrac12(n-1)\,\frac{\alpha}{4\pi}\,\frac1{r^4}$.
We therefore define the constant $Z_F$ as
\ba
    Z_F = 
    \lim\limits_{r\to\infty} \frac{r^4 F(r)}{ \tfrac12(n-1)\,\frac{\alpha}{4\pi}}
    \;. \nonumber
\ea 
With $Z_F$ defined in this way, the integral is finite.
We thus define our regularization prescription to be
\ba\label{Eq:reg-method}
    I_n = \int d^3r\,r^n F(r)_{\rm reg}
    = \int d^3r\,r^n \biggl[F(r) 
    + Z_F\Bigl\{\tfrac23\,n\,s(r)+(n+3)\,p(r)\Bigr\}\biggr] 
    = \mbox{finite}.
\ea
This regularization method has the advantage that it can be applied 
to any EMT property. Applying this method to the $D$-term where 
$n=2$ and $F(r) = p(r)$ with $Z_F = \frac13$ or $F(r) = -\,\frac4{15}\,s(r)$ 
with $Z_F = \frac8{15}$ we obtain
\ba
    D_{\rm prot, reg} = \begin{cases}  
        \hspace{6.8mm}
        M \int d^3 r\; r^2 p(r)_{\rm reg} 
      = M \int d^3 r\; r^2 \biggl[
        \hspace{6.8mm}
        p(r) + \;\frac13\,\Bigl\{\tfrac43\,s(r)+5\,p(r)\Bigr\}\biggr] \,, \\  
    - \frac{4}{15}\,M \int d^3 r\; r^2 s(r)_{\rm reg}
    = M \int d^3 r\; r^2 \biggl[- \frac{4}{15} \,s(r) 
    +\frac8{15}\Bigl\{\tfrac43\,s(r)+5\,p(r)\Bigr\}\biggr] \, ,\phantom{\Biggl|}
    \end{cases} \nonumber
\ea
which reproduces, in either case, unambiguously the result (\ref{Eq:D-prot-reg})
derived in \cite{Varma:2020crx}. Numerically, we obtain 
\ba
    D_{\rm neut\phantom{,reg}}     &=& -0.317\,(\hbar c)^2  , \\
    D_{\rm prot,\rm reg}          &=& -0.322\,(\hbar c)^2  .
\ea
Our result for the neutron confirms that the concept of proton $D$-term 
regularization proposed in \cite{Varma:2020crx} makes physically sense because 
one expects proton and neutron to have very similar properties if one disregards 
isospin violating effects and ``neglects'' electromagnetic effects (with the
neglect of electromagnetic effects defined by the above regularization). 

Before ending this section, let us briefly explain the dashed dotted line 
in Fig.~\ref{fig3-A-D-FFs}b which motivated in Ref.~\cite{Varma:2020crx} the
idea to introduce a regularized value $D_{\rm prot,reg}$. If we refrain from 
small $(-t)$ where QED effects play a role and large $(-t)$ where the model
is not applicable, i.e.\ if we focus on an intermediate range 
$0.1\,{\rm GeV}^2\lesssim (-t) \lesssim 0.5\,{\rm GeV}^2$, 
then the proton $D(t)$ can be well approximated by 
$D(t) \approx D_{\rm prot,reg}/(1-t/m_D^2)^n$ with $m_D = 0.985\,\rm GeV$ and $n=3$
\cite{Varma:2020crx}.
In this $(-t)$-range, this approximation and the numerical proton and
neutron results for $D(t)$ can all be hardly distinguished on the scale of 
Fig.~\ref{fig3-A-D-FFs}b. Thus, $D_{\rm prot,reg}$ is useful to provide an
approximation to the proton $D(t)$ in an intermediate $(-t)$-range.
This is an indication for the practical usefulness of the concept of a
regularized proton $D$-term. We will see a stronger argument in
Sec.~\ref{Sec-8:compare-to-lattice}, after making further use of
the new regularization method.

\section{EMT radii and neutron size}
\label{Sec-7:radii}

In Sec.~\ref{Sec-5:proton-vs-neutron} we saw that the dust distribution in the neutron
occupies a smaller volume as compared to the proton with $R_n = 1.04\,{\rm fm}$ vs
$R_p = 1.05\,{\rm fm}$. Within the model, we can use the dust mean square root radii
for a somewhat more quantitative comparison of the sizes of neutron and proton which
yields
\be\label{Eq:dust-radius}
    \la r^2_{\rm dust}\ra{ }^{1/2}_n = 0.704\,{\rm fm}, \quad
    \la r^2_{\rm dust}\ra{ }^{1/2}_p = 0.710\,{\rm fm}.
\ee
In the proton case, the dust radius corresponds to a physical quantity as it defines
its charge radius and hence the ``proton size''. In the neutron case, there is no such
association. The neutron mean square charge radius is with 
$\la r^2_{\rm ch}\ra^{ }_n = (-0.1155 \pm 0.0017)\,\rm fm^2$ \cite{ParticleDataGroup:2024cfk}
negative and tells us nothing about the ``neutron size,'' albeit it gives insights about
the electric charge distribution in the neutron (whose description is also beyond the
present model, see Sec.~\ref{Sec-4:model-neutron}).

In order to meaningfully compare the sizes of proton and neutron, we can make use of the
EMT radii in Eq.~(\ref{Eq:EMT-radii}). The latter are well-defined for the neutron but  
diverge for the proton due to the large-$r$ behavior of the EMT distributions in
Table~\ref{tab1}. In this work, based on the new regularization method 
in Sec.~\ref{Sec-6:EMT-FFs-and-Dterm}, we are in the position to compute these 
quantities for the proton and compare to the corresponding neutron results.
To regularize the proton EMT radii, we use Eq.~(\ref{Eq:reg-method}) with $n=2$ and
respectively $F(r)=\frac23\,s(r)+p(r)$ with $Z_F = -1$ for the mechanical radius
and $F(r)=T_{00}(r)$ with $Z_F = 1$ for the mass radius. The neutron and proton
results are 
\ba\label{Eq:mech-radius-n-p}
    \la r^2_{\rm mech}\ra^{1/2}_{\rm neut\phantom{,\rm reg}} = 0.769\,{\rm fm}\,, && 
    \la r^2_{\rm mech}\ra^{1/2}_{\rm prot,\rm reg} =  0.797\,{\rm fm}\,, \\
\label{Eq:mass-radius-n-p}
    \la r^2_{\rm mass}\ra^{1/2}_{\rm neut\phantom{,\rm reg}} =  0.755\,{\rm fm}\,, &&
    \la r^2_{\rm mass}\ra^{1/2}_{\rm prot,\rm reg}          = 0.759\,{\rm fm}\,.
\ea
Several comments are in order. We recall that the dust radius
(\ref{Eq:dust-radius}) has a physical meaning (in terms of the charge radius)
only for the proton, while the EMT radii
(\ref{Eq:mech-radius-n-p},~\ref{Eq:mass-radius-n-p}) are in principle both
observable providing a path to a meaningful comparison of the proton and
neutron sizes. All neutron and proton radii are numerically very similar,
see Eqs.~(\ref{Eq:dust-radius}--\ref{Eq:mass-radius-n-p}).
However, in all cases the neutron is slightly smaller than the proton.
Rather than the absolute numbers in
Eqs.~(\ref{Eq:dust-radius}-\ref{Eq:mass-radius-n-p}), a more robust
model prediction is that the mechanical radii of neutron and proton
are respectively $8\,\%$ and $12\,\%$ larger than the proton charge
radius. For both particles, the mass radius is (rounded off) about
$6\,\%$ larger than the proton charge radius. 

The fact that the neutron appears systematically smaller than the proton
across the various radii (\ref{Eq:dust-radius}-\ref{Eq:mass-radius-n-p})
can be intuitively understood: the repulsive electrostatic forces present
in the proton (but absent in the neutron) tend to spread out the matter distribution
in the proton. The effect is relatively small because in the model the
electromagnetic forces are three orders of magnitude weaker than strong forces.
We recall that the strong forces in the model are the relatively weak residual
forces. In nature, the QCD color forces are far stronger such that experimentally
one should expect even smaller differences between the EMT radii of proton and neutron.

Noteworthy is that the proton radii in 
Eqs.~(\ref{Eq:mech-radius-n-p}) and (\ref{Eq:mech-radius-n-p}) are the results of a
regularization which corresponds to subtracting off infinities. It is a remarkable
and non-trivial result that our regularization preserves the expected physical
feature that the proton is larger (and only slightly larger) than the neutron.

\section{Simulating the physical situation for proton and neutron}
\label{Sec-8:compare-to-lattice}

An interesting question is whether the divergence for $t\to 0$ of the $D(t)$
of proton can be observed experimentally. The model can be used to shed light 
on this question but for that one first needs to make the model description more 
realistic. For that we will address two aspects.
(i) In the model the proton charge radius is about $15\,\%$ 
smaller than the experimental value. (ii)
The neutron and regularized proton $D$-terms are about an order
of magnitude smaller than, e.g., the $D$-terms in the more realistic 
chiral quark soliton model \cite{Goeke:2007fp} or in lattice QCD 
\cite{Hackett:2023rif}.

In order to address these two aspects we
introduce two free model parameters $\lambda_m$ and $\lambda_g$ 
which simultaneously rescale respectively the masses $m_i$ and 
coupling constants $g_i$ of the scalar and vector mesons 
($i = $ {\footnotesize $S$, $V$}) as follows
\be\label{Eq:resc}
    m_i \to m_i^\prime = \lambda_m \, m_i \, , \quad 
    g_i \to g_i^\prime = \lambda_g \, g_i \,.
\ee
The neutron stress tensor properties scale homogeneously with
$\lambda_m$ and $\lambda_g$ (in a specific way depending on the property).
In the proton case, the stress tensor properties scale homogeneously with
$\lambda_m$ (because all the relevant dimensionful parameters are rescaled
including the photon mass, $m_\gamma^\prime = \lambda_m\,m_\gamma$, which happens
to be zero) but not with~$\lambda_g$ (because in Eq.~(\ref{Eq:resc}) we do not
simultaneously rescale the electromagnetic coupling constant, keeping it fixed). 
Due to the smallness of the electromagnetic coupling constant there is,
however, an approximate scaling with $\lambda_g$.\footnote{The energy density
   has no definite scaling behavior
   because the model parameter $m_{\rm dust}$ is adjusted to reproduce the proton
   mass according to footnote~\ref{footnote-parameter-fixing} as we vary the $\lambda_i$. 
   As a consequence of that, the energy density can turn negative in the vicinity of $r=0$ 
   for $\lambda_g \ge 2.1$ when $\lambda_m = 0.85$. This may be considered a ``finite-size 
   renormalization artifact'' in an interacting theory as coupling constants are varied. 
   In practice, renormalization effects do not always preserve all physical properties, 
   like positivity of energy density in this case. As $m_{\rm dust}$ enters only in 
   combination with $\rho(r)$ (normalized to unity) in $T_{00}(r)$, it is just an additive
   constant in the nucleon mass and does not affect mechanical stability and properties 
   related to the stress tensor. }

In this way, in the model the nucleon properties become functions of 
$\lambda_m$ and $\lambda_g$. As can be seen from the Yukawa tails of
the meson fields proportional to $\exp(-m_i r)$, 
Eqs.~(\ref{eq:gsphi},~\ref{eq:gswzero},~\ref{neut-eq:gsphi},~\ref{neut-eq:gswzero}),
one can think of the change $m_i \to m_i^\prime$ in Eq.~(\ref{Eq:resc}) as
implying a redefinition of the unit of length.
As a consequence, for instance the mean square square root dust radius
(which corresponds to the proton charge radius)
depends on $\lambda_m$ as
\be\label{Eq:scaling-dust-radius}
    \la r_{\rm dust}^2(\lambda_m,\lambda_g)\ra^{1/2} 
    = \lambda_m^{-1} \la r_{\rm dust}^2\ra^{1/2}\,.
\ee
Here and in the following, quantities written without explicit
dependence on the $\lambda_i$ refer to unscaled model results.
For instance, in Eq.~(\ref{Eq:scaling-dust-radius})
$\la r_{\rm dust}^2\ra=\la r_{\rm dust}^2(\lambda_m,\lambda_g)\ra|_{\lambda_m=\lambda_g=1}$.
  
Setting $\lambda_m = 0.85$ yields a proton charge radius of
$0.83\,\rm fm$ in agreement with the experimental result 
inferred from elastic electron-proton scattering experiments 
$(0.831 \pm 0.007_{\rm stat} \pm 0.012_{\rm syst})\,\rm fm$
\cite{Xiong:2019umf}. For this value of $\lambda_m$, 
the electromagnetic proton-neutron mass difference which scales as
$(M_{\rm prot}-M_{\rm neut})(\lambda_m,\lambda_g)_{\rm e.m.}=\lambda_m\,(M_{\rm prot}-M_{\rm neut})_{e.m.}$,
is reduced by $15\,\%$ which is  within the theoretical
uncertainties of the lattice result quoted in
Eq.~(\ref{Eq:mass-comparison}). We conclude that a rescaling
with $\lambda_m=0.85$ yields a realistic description of the
electromagnetic properties. Let us recall that the proton
mass is always exactly described according to the parameter fixing
adopted in this work, see footnote~\ref{footnote-parameter-fixing}.

A rescaling with $\lambda_g>1$ increases the strength of the strong
model forces which for $\lambda_g=1$ correspond to residual
nuclear forces. To obtain a realistic result for the
$D$-term, the residual nuclear forces need to be scaled up to
more adequately simulate the QCD forces. The pressure (similarly
the shear force) and the $D(t)$ form factor behave as
\ba
    p(r,\lambda_m,\lambda_g) 
    &=& \lambda_m^4 \, \lambda_g^2 \; p(r\lambda_m^{-1})_{\rm strong}^{\phantom{0}} 
      + \lambda_m^4 \;               p(r\lambda_m^{-1})_{\rm em}^{\phantom{0}} \,,
    \phantom{\biggl|}\\
    D(t,\lambda_m,\lambda_g) 
    &=& \lambda_m^{-1} \lambda_g^2 \, D(t\lambda_m^2)_{\rm strong}^{\phantom{0}} 
      + \lambda_m^{-1} \, D(t\lambda_m^2)_{\rm em}^{\phantom{0}} \,.
\ea
To determine the value of $\lambda_g$ needed to scale up the strong model
forces to obtain a $D$-term as large as determined in lattice QCD
we use as guideline the results from \cite{Hackett:2023rif} obtained from 
gauge field configurations generated on a $L^3 \times T = (48a)^3 \times (96a)$
lattice \cite{Edwards-unpublished-2016} based on  L\"uscher-Weisz gauge
action \cite{Luscher:1984xn} with $N_f= 2 + 1$ flavors of clover-improved
Wilson fermions \cite{Sheikholeslami:1985ij}, lattice spacing 
$a \approx 0.091\,\rm fm$, and nearly physical pion masses 
$m_\pi \approx 170\,\rm MeV$~\cite{Park:2021ypf}.

\begin{figure}[t!]
  \begin{center}
    \includegraphics[height=3.7cm]{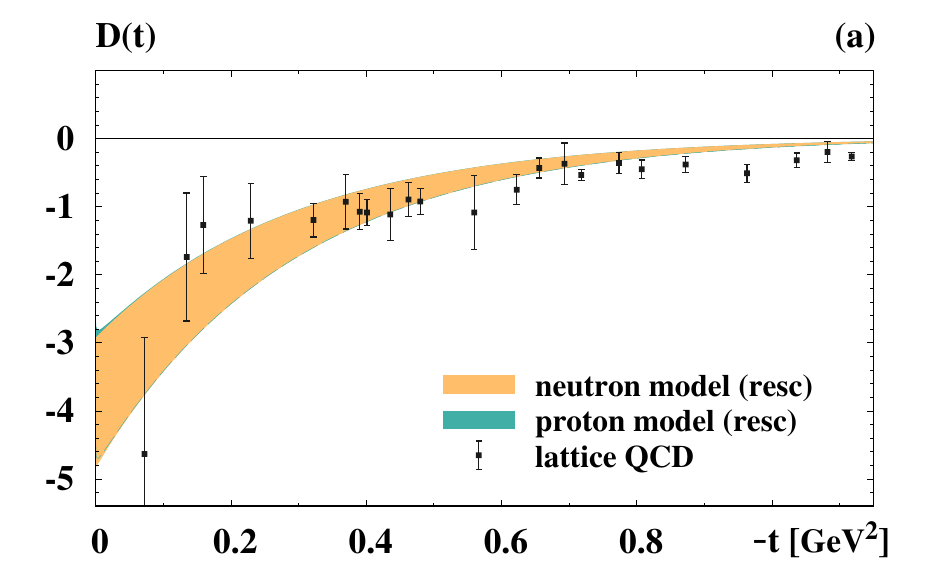}%
    \includegraphics[height=3.7cm]{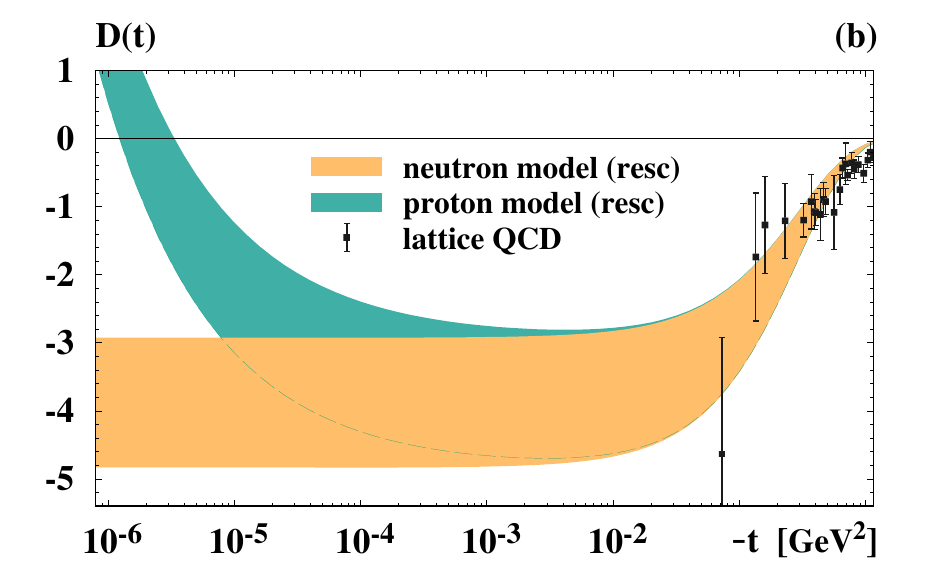}%
    \includegraphics[height=3.7cm]{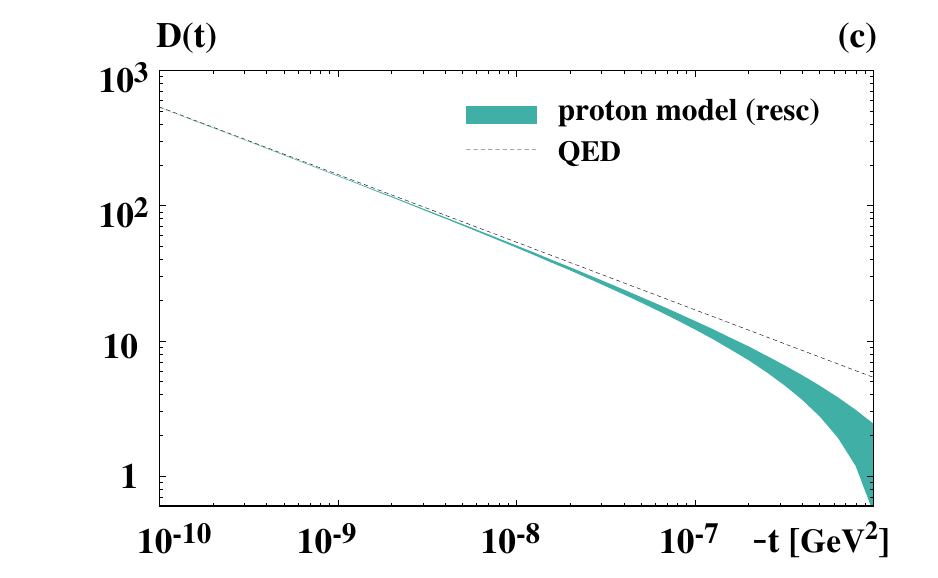}%
  \end{center}
  \caption{\label{fig4-rescaled}
    (a) $D(t)$ of neutron and proton in the model with strong 
    interaction parameters rescaled according to Eq.~(\ref{Eq:resc})
    by $\lambda_m = 0.85$ which yields the correct proton charge radius 
    and $\lambda_g=3.2\pm 0.4$ which yields the displayed uncertainty band
    and a good description of the lattice data from Ref.~\cite{Hackett:2023rif}
    at $(-t)\lesssim 1\,{\rm GeV}^2$ that are shown in the figure for comparison. 
    (b) The same as panel (a) but on a logarithmic $(-t)$-scale illustrating that 
    the difference between $D(t)$ of neutron and proton becomes noticeable only 
    for $(-t)\lesssim 10^{-4}\,\rm GeV^2$. (c) $D(t)$ of proton 
    $(-t)\lesssim 10^{-6}\,\rm GeV^2$ in comparison to the QED prediction.}
\end{figure}

In the lattice study \cite{Hackett:2023rif} results for the total
(quark + gluon) nucleon $D(t)$ form factor were presented which is
renormalization scale independent allowing one to carry out a
direct comparison to a low-energy effective theory such as our model.
In the lattice calculation \cite{Hackett:2023rif} isospin breaking and
electromagnetic effects were not considered. This makes the proton
as studied in  \cite{Hackett:2023rif} electrically neutral and
mass-degenerate with the neutron.
The $D$-term appears in the decomposition of EMT matrix elements with two
powers of $\Delta^\mu$ in Eq.~(\ref{Eq:EMT-FFs-def}) and cannot be computed
on the lattice at zero momentum transfer. The smallest available value 
in \cite{Hackett:2023rif} was $(-t)=0.067\,\rm GeV^2$. The extrapolation
to $t=0$ can be obtained by different methods with similar results.
For definiteness, we use here the dipole fit result which has a somewhat
larger uncertainty and hence provides a more conservative estimate for the
total nucleon $D$-term \cite{Hackett:2023rif}
\be
   D_{\rm lattice} = -(3.87\pm 0.97)\,.
\ee
Recall that this corresponds to an ``uncharged proton'' or neutron
in the absence of isospin violating effects.
In the model this value is reproduced well for ($\lambda_m=0.85$ and)
$\lambda_g = (3.2\pm 0.4)$ which yields respectively
\ba
   D_{\rm neut\phantom{,reg}} &=& -(3.88\pm 0.96), \\
   D_{\rm prot, reg}        &=&  -(3.89\pm 0.96).
\ea
Remarkably, the neutron and regularized proton model results agree within
1\permil\ accuracy. With $\lambda_m$ and $\lambda_g$ fixed in the above-explained
way, we obtain the results for $D(t)$ shown in Fig.~\ref{fig4-rescaled}.

Fig.~\ref{fig4-rescaled}a shows that the model results for proton and neutron
practically overlap. Although we adjusted the parameter $\lambda_g$
to reproduce only one single value of $D(t)$ --- namely at $t=0$, i.e.\ the
$D$-term --- we nevertheless observe an excellent agreement with the lattice data
\cite{Hackett:2023rif} over an extended $t$-range in Fig.~\ref{fig4-rescaled}a.
The $\chi^2$ per degree of freedom, $\chi_{d.o.f.}^2$, of the model description of
lattice QCD data is $\chi_{d.o.f.}^2 = 1.05$ in the region $(-t) < 0.8\,\rm GeV^2$.
Beyond that we observe a tendency of the model to undershoot the lattice data.
We find worse $\chi_{d.o.f.}^2 = 1.27$ for $(-t) < 0.9\,\rm GeV^2$,
and $\chi_{d.o.f.}^2 = 1.65$ for $(-t) < 1\,\rm GeV^2$, although in view of the 
simplicity of the model, even the last value is acceptable.

Notice that we adjusted $\lambda_g$ to describe $D(t)$ at $t=0$ and made no effort
to optimize the $\chi^2$ over a range of $t$ because mean field approaches to form
factor computations are expected to become sensitive to relativistic recoil with
increasing $(-t)$ with a range of applicability up to $(-t) \approx 1\,\rm GeV^2$
\cite{Braaten:1986iw,Ji:1991ff,Christov:1995vm}. 
The description of asymptotically large $(-t)$ is in the regime of perturbative QCD
\cite{Tong:2021ctu,Tong:2022zax} and certainly beyond the scope of effective low energy
models. For the following, we conclude that the model gives a very good description of
the lattice data on $D(t)$ \cite{Hackett:2023rif} up to $(-t) \lesssim 1\,\rm GeV^2$.
It is a remarkable and nontrivial outcome that by fixing 1 parameter,
$\lambda_g$, the model is able to describe ${\cal O}(20)$ lattice data points in
Fig.~\ref{fig4-rescaled}a (after fixing $\lambda_m$ by means of the proton charge
radius).

Fig.~\ref{fig4-rescaled}b shows the same result as Fig.~\ref{fig4-rescaled}a but
with a logarithmic scale for the $(-t)$--axis to illustrate when one may expect to
start seeing deviations. Until $(-t) \approx 10^{-4}\,\rm GeV^2$ there is no appreciable
difference between the two curves. In other words, until then the proton and neutron
form factors $D(t)$ may be expected to look basically the same. 
A noticeable difference between the $D(t)$ of these two baryons becomes clearly visible
only below  $(-t) \approx 10^{-5}\,\rm GeV^2$. At still smaller $(-t)$, somewhere in the
range $10^{-6}\,\rm GeV^2 < (-t) <  10^{-5}\,\rm GeV^2$ depending on the parameter $\lambda_g$,
the proton $D(t)$ changes sign and becomes positive.

Fig.~\ref{fig4-rescaled}c shows the proton $D(t)$ in the region $(-t) \le 10^{-6}\rm GeV^2$
where it is positive. The uncertainty band due to $\lambda_g = (3.2\pm 0.4)$ shrinks visibly
as we enter the regime of small $(-t)$ dominated by QED effects. The proton result
starts to approach the QED result which is universal for all charged particles, see
Eq.~(\ref{eq:D-term-QED}), for $(-t) \lesssim 10^{-7}\rm GeV^2$. On the scale of the
double-logarithmic plot in Fig.~\ref{fig4-rescaled}c, the model result for the proton
$D(t)$ and the QED result are practically indistinguishable in the region of 
$10^{-10}\,\rm GeV^2 \le (-t) \lesssim 10^{-8}\,\rm GeV^2$. At still smaller $(-t)$,
our computational method becomes numerically unstable. However, the asymptotic QED
result in Eq.~(\ref{eq:D-term-QED}) can be derived analytically.

For that we use  $s(r) = c_p \,\frac1{r^4}+\dots$ for the shear force at $r\gg R_p$
from Table~\ref{tab1} where $c_p = \frac{\alpha}{4\pi}$ and the parenthesis indicates
exponentially decaying terms. With this result, we determine $\tilde{D}(r)$ from
Eq.~(\ref{Eq:def-s-p}) which yields at $r \gg R_p$ 
\be\label{Eq:Dtilde-asymp}
    \tilde{D}(r) = \frac{c_p M}{2r^2} + \dots 
\ee
where the dots indicate exponentially suppressed terms. Now we can invert the
Fourier transform in Eq.~(\ref{Eq:def-s-p}) by separating $\tilde{D}(r)$ in two
contributions: one due to strong fields at all $r$ plus the electromagnetic
part at $r<R_p$, and the other one due to the electromagnetic part in
Eq.~(\ref{Eq:Dtilde-asymp}) at $r > R_p$.
The former yields after the Fourier transform a finite contribution at $t=0$.
The latter yields after integrating out the angular variables in spherical coordinates
\be
   D(t)
   = \int d^3r\;e^{i\vec{\Delta}\cdot\vec{r}}\,\tilde{D}(r)
   = 2\pi\,c_p M \int\limits_{R_p}^\infty dr\;\frac{\sin(|\vec{\Delta}|r)}{|\vec{\Delta}|r} +\mbox{finite}
   = 2\pi\,c_p M \,\frac{1}{\sqrt{-t}}\int\limits_{R_p\sqrt{-t}}^\infty dy\;\frac{\sin y}{y} +\mbox{finite}
\ee
where we substituted $y=|\vec{\Delta}|r$ and used $t=-\vec{\Delta}^2$
or $|\vec{\Delta}|=\sqrt{-t}$. In the limit of very small $(-t)$ when
$\epsilon = R_p\sqrt{-t}\ll 1$, the integral yields
$\frac{\pi}{2}+\epsilon+{\cal O}(\epsilon^3)$. The $\epsilon$-terms
can be combined with the finite terms. Thus, we obtain
\be
   D(t) = \frac{\pi^2\,c_p M}{\sqrt{-t}} + \mbox{finite}
   = \alpha\,\frac{\pi M}{4\sqrt{-t}} + \mbox{finite}
\ee
which agrees with the QED result in Eq.~(\ref{eq:D-term-QED}).

Thus, the model exactly reproduces the QED prediction in the limit $t\to0$
and lattice QCD in the region for $0.06\,\rm GeV^2 < t \lesssim 1\,\rm GeV^2$
which gives confidence that it can reasonably interpolate between what we know
from QED and lattice QCD about the $D(t)$ form factor of the proton.
Based on our model results, we can draw two conclusions. First, the
proton and neutron $D(t)$ can be expected to be practically the same
down to $(-t)\approx 10^{-4}\,\rm GeV^2$. Second, it does make sense
to introduce the notion of a regularized proton $D$-term to approximate
the exact proton $D(t)\approx D_{\rm prot,reg}/(1-t/m_D^2)^n$ with
some effective parameter $n$, for instance, $n=2$ as used in the
lattice data fit in Ref.~\cite{Hackett:2023rif} or $n=3$ as used
in \cite{Varma:2020crx} or in the first attempt to extract the
proton $D(t)$ from data \cite{Nature}. 

\newpage
\section{Conclusions}
\label{Sec-9:conclusions}

The EMT form factor $D(t)$ of neutral hadrons is negative and finite at $t=0$,
while that of charged hadrons changes sign at small values of $(-t)$ and exhibits
for $t\to 0$ a divergence $D(t) \propto 1/\sqrt{-t}$ as first observed for
charged pions in chiral perturbation theory \cite{Kubis:1999db} and for proton
\cite{Varma:2020crx} in the model by Bia\l ynicki-Birula \cite{BialynickiBirula:1993ce}.
The divergence of $D(t)$ is a universal effect for any charged particle due to the
masslessness of the photon \cite{Donoghue:2001qc,Metz:2021lqv,Freese:2022jlu}.
In this work, we investigated the important question whether this effect can be
observed experimentally for the proton, and whether protons and neutrons must be
treated differently in phenomenological studies of hard exclusive reactions.

To address this question, we constructed a classical model of the neutron 
which differs from the original proton model
\cite{BialynickiBirula:1993ce,Varma:2020crx} only by making the dust particles, which
constitute the matter in that model, electrically neutral. Despite its simplicity,
this simple model accurately accounts for the electromagnetic proton-neutron mass
difference. For the neutron we obtain a negative and, for all $(-t)$-values including
$t=0$, finite $D(t)$ form factor which is numerically very close to the regularized
proton $D$-term proposed in Ref.~\cite{Varma:2020crx} arguing that the QED divergence
of $D(t)$ might most likely be out of reach for experimental detection such that
the proton $D(t)$ would practically look finite.

To examine this assertion quantitatively it was necessary to alter the original
proton model formulation, where the proton charge radius is about $15\,\%$ too small
(though ``not completely out of touch with reality'' \cite{BialynickiBirula:1993ce})
and the $D$-term is about an order of magnitude smaller \cite{Varma:2020crx} compared
to results from more realistic hadronic models and lattice QCD (because the residual
nuclear forces underlying the model are far weaker than the strong forces between quarks).
To make the modelling of the strong forces more realistic, we introduced two model
parameters: one to bring the model result for the proton charge radius in agreement
with experiment, and the other to bring the model result for the $D$-term in agreement
with lattice data \cite{Hackett:2023rif}. Remarkably, this resulted in a very
good description of lattice QCD data on $D(t)$ up to $(-t) \lesssim {\rm 1\,GeV^2}$.
We have also analytically proven that the model exactly reproduces the QED
divergence of $D(t)$ for the proton. 

Equipped with this realistic model incorporating all that is currently known
about $D(t)$ from QED and QCD from $t\to 0$ until $(-t) \lesssim {\rm 1\,GeV^2}$,
we have shown that one would need to go to $(-t)$ below $10^{-4}\,\rm GeV^2$ to
observe deviations between the proton and neutron $D(t)$ form factors.
In order to verify experimentally the QED prediction for the proton
$D(t) = \alpha\,\pi\, M/(4\sqrt{-t})+\dots$ one would need to reach $(-t)$
values of  $10^{-8}\,\rm GeV^2$ or smaller.
In the very first attempt to extract information related to $D(t)$ from Jefferson
Lab data the minimal $(-t)$ value was $(-t)_{\rm min}=0.09\,\rm GeV^2$ \cite{Nature}.
Impact studies for the planned Electron-Ion Collider assume $(-t_{\rm min}) = 0.03\,{\rm GeV}^2$
\cite{Aschenauer:2013hhw,AbdulKhalek:2021gbh}.
Our study therefore leads to the conclusion that, in the foreseeable future, the
experimental observation of QED effects in the proton $D(t)$ will be out of reach.
This means that, for all practical purposes in phenomenological studies of hard
exclusive reactions, the proton and neutron $D(t)$ will look the same: negative and
finite down to the lowest measurable $(-t)$ values in ongoing and planned experiments.
This implies that it is a reasonable step to introduce the notion of a
``regularized proton $D$-term'' $D_{\rm prot,reg}$ for Ans\"atze of the type
$D(t)\approx D_{\rm prot,reg}/(1-t/m_D^2)^n$ for the proton for $(-t)> 10^{-4}\,\rm GeV^2$
in phenomenological applications which can be expected to be very similar to the
neutron $D$-term. In fact, in the model $D_{\rm neut}$ and $D_{\rm prot,reg}$ coincide
within 1$\,$\permil\ accuracy.

We generalized the method to regularize the proton $D$-term of Ref.~\cite{Varma:2020crx}
to other EMT properties, and explored it to compute the EMT radii of proton which are
divergent (similarly to the $D$-term) due to the long-range Coulomb field contribution.
Comparing the mechanical radius and the mass radius of the proton to those of the neutron
(which are finite), we observed that the neutron is systematically slightly smaller than
the proton because latter appears slightly swollen compared to the neutron, due to Coulomb
repulsion. As a by-product of this study, we point out that the EMT radii allow, in contrast to the mean square charge radius, to characterize the neutron size.
According to the model, proton and neutron will appear as nearly the same size
--- as one would intuitively expect for these SU(2) partners which are very
similar in so many respects, if one refrains from electromagnetic corrections
and isospin breaking effects. Based on measurements of the hard exclusive reactions,
it will be possible to test this prediction experimentally.

\ \\
\noindent{\bf Acknowledgments.} 
The authors are indebted to Phiala Shannahan for sharing the lattice results from 
Ref.~\cite{Hackett:2023rif}. This work was supported by  NSF under the Award
No.\ 2412625, and DOE under the umbrella of the Quark-Gluon Tomography (QGT)
Topical Collaboration with Award No.\ DE-SC0023646.

\end{document}